   \documentclass[prd,twocolumn,superscriptaddress,nofootinbib,showpacs]{revtex4}

\usepackage{graphicx}
\usepackage{dcolumn}
\usepackage{bm}
\begin{document}

\title{Is the cold spot responsible for the CMB North-South asymmetry?} 

\author{Armando Bernui}

\affiliation{Divis\~ao de Astrof\'{\i}sica, Instituto Nacional de Pesquisas Espaciais \\ 
Av. dos Astronautas 1758, 12227-010, S\~ao Jos\'e dos Campos, SP, Brazil}
\affiliation{Centro Brasileiro de Pesquisas F\'{\i}sicas, \\
Rua Dr. Xavier Sigaud 150, 22290-180, Rio de Janeiro, RJ, Brazil}


\begin{abstract}
Several intriguing phenomena, unlikely within the standard inflationary cosmology, were 
reported in the cosmic microwave background (CMB) data from WMAP and appear to be 
uncorrelated. 
Two of these phenomena, termed CMB anomalies, are representative of their disparate 
nature: 
the North-South asymmetry in the CMB angular-correlation strength, inconsistent with an 
isotropic universe, 
and the cold spot, producing a significant deviation from Gaussianity. 
We find a correlation between them, at medium angular scales ($\ell = 11 - 20$): 
we show that a successive diminution of the cold spot (absolute-value) temperature implies 
a monotonic decrease of the North-South asymmetry power, and moreover we 
find that the cold spot supplies 60\% of such power. 
\end{abstract}

\pacs{98.70.Vc,98.80.Es}

\maketitle

\section{Introduction} \label{Introduction} 

Successive data releases from the Wilkinson Microwave Anisotropy Probe 
(WMAP)~\cite{wmap} have confirmed the validity of the standard inflationary cosmological 
model $\Lambda$CDM, which asserts that the observed cosmic microwave background 
(CMB) temperature f\/luctuations are a stochastic realization of an isotropic Gaussian random 
field on the celestial sphere. 
This means that the CMB sky should exhibit statistical isotropy and statistical Gaussianity 
attributes. 

Close scrutiny of the CMB WMAP maps have revealed highly significant departures 
both from statistical isotropy and from Gaussianity at large and medium angular scales. 
Evidences of an anomalous power asymmetry of the CMB angular correlations between the 
northern and southern ecliptic hemispheres (termed the NS-asymmetry) indicate 
that the CMB temperature field is inconsistent with the statistical isotropy expected in the 
model~\cite{NS-asymmetry}. 
Another intriguing detection concerns an anomalously cold and large spot (termed the 
cold spot), centered at $(l,b) \simeq (209^{\circ},-57^{\circ})$ in galactic coordinates, 
which causes a significant deviation from Gaussianity~\cite{cold-spot}. 

Many attempts have been made to explain these phenomena, termed CMB anomalies, 
in particular, within the standard inf\/lationary $\Lambda$CDM scenario 
where they are unlikely at $\lesssim 1 \%$ of probability.
Hypotheses like 
instrumental noise, 
systematic errors (e.g., in the mapmaking process), 
the inhomogeneous exposure function of the probe, 
incomplete sky-data (due to the cut-sky mask), 
unmodeled foreground emissions~\cite{foregrounds}, 
and physical mechanisms that break statistical isotropy 
(e.g., during the epoch of inflation or during the decoupling era, as for instance, 
primordial magnetic fields)~\cite{mechanisms}, 
have been extensively investigated. 
Unfortunately, none of them seems to explain satisfactorily the reported CMB anomalies. 

Another way to comprehend these phenomena is to look for possible correlations 
between them, or \mbox{establish} their absence. 
Clearly, finding a cause-effect relationship between CMB anomalies would simplify 
the search for their origin. 
For instance,~\cite{correl-anom1} proved the absence of correlation between the 
alignment of low order multipoles and the observed lack of CMB angular correlations on 
scales $> 60^{\circ}$. 
On the other hand,~\cite{correl-anom2} found that the alignment of the CMB 
quadrupole and octopole is not responsible for the anomalous NS-asymmetry at large 
angles. 
Likewise, using needlets~\cite{correl-anom3} detected anomalous spots in the needlets' 
coefficients map. 
However, it seems possible that one of such spots could be caused by the CMB cold spot. 
In such a case the hemispherical asymmetry found in the needlets' power 
spectrum~\cite{correl-anom3} could be related to the CMB cold spot. 
 
In this work we show that the cold spot is responsible for 60\% of the NS-asymmetry 
strength in WMAP maps at medium angular scales (i.e., maps with multipoles $\ell = 11 - 20$). 
Firstly, we prove that the NS-asymmetry is present at 94\%$-$98\% C.L. in these maps. 
Then we show that gradually reducing the cold-spot temperature (turning it less cold) 
turns the NS-asymmetry effect less and less statistically significant. 
When the cold spot is suppressed the statistical significance of the NS-asymmetry 
phenomenon goes to the level found in the average from statistically-isotropic Gaussian 
CMB maps.

\section{WMAP data} \label{WMAP data} 

Substantial ef\/forts done by the WMAP science team to minimize foregrounds and to limit 
systematic errors resulted in the five-year foreground-reduced single-frequency Q, V, and 
W band-maps~\cite{wmap}. 
Our aim is to investigate the angular distribution and statistical features of the CMB 
temperature field in these Q, V, and W maps at medium angular scales, that is, maps with 
multipole components $\ell = 11 - 20$. 
To obtain such data we first perform the multipolar decomposition of the original maps, 
applying the recommended KQ75 mask~\cite{wmap}, by using the {\sc anafast} 
code~\cite{Gorski}. 
After that, we select the multipole components $11 \leq \ell \leq 20$ and generate the 
corresponding CMB map with the {\sc synfast} code~\cite{Gorski}. 

\begin{figure} 
\hspace{0.3cm}
\includegraphics[width=7.5cm,height=13cm]{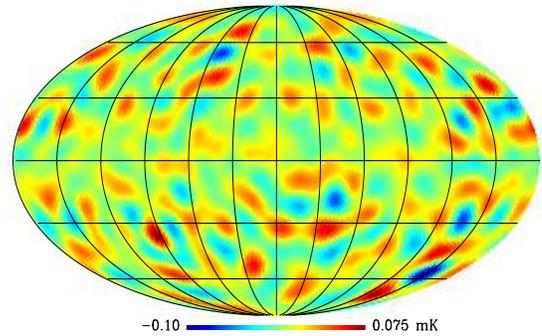}

\vspace{-7.3cm}
\mbox{\hspace{-0.5cm}
\includegraphics[width=9cm,height=13.5cm]{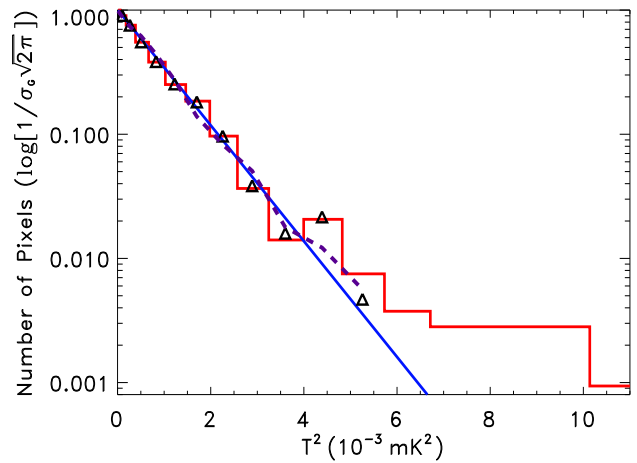}}
\vspace{-6.5cm}
\caption{\label{fig1} 
Top: 
This is the V map with multipole components 
$\,\ell = 11 - 20$, obtained using the KQ75 mask. 
Bottom: 
This plot shows the temperature distribution of this V map, 
considering data outside the KQ75 region, where: 
(i) the straight (blue) line is the expected Gaussian distribution with variance 
$\sigma_{\mbox{\sc\footnotesize g}}^2 = 4.65 \times 10^{-4}$ mK$^{2}$, 
(ii) the histogram (red) curve corresponds to the negative pixel's temperatures, 
(iii) the dashed (violet) line corresponds to the positive pixel's temperatures, and 
(iv) the triangles represent the distribution of the negative pixels' temperatures 
corresponding to a modified V map where the cold-spot temperature is 70\% of its 
original value. 
} 
\end{figure}

In the top panel of Fig.~\ref{fig1} we exhibit the V map containing the multipoles 
$\ell = 11 - 20$. 
The Q and W maps are fully similar to this V map, as corroborated by Pearson's 
coefficient $p_{_{\mbox{\sc\footnotesize ab}}}$ that measures the pixel-to-pixel correlation 
between A and B maps: $p_{_{\mbox{\sc\footnotesize vq}}}=0.9985$, 
$p_{_{\mbox{\sc\footnotesize vw}}}=0.9982$, and $p_{_{\mbox{\sc\footnotesize qw}}}=0.9976$. 
In this V map one clearly observes a large blue spot in the lower right corner: 
this is the cold spot with radius about $8^{\circ}$ and centered at 
$(l,b) \!\simeq\! (209^{\circ},-57^{\circ})$. 

A statistical analysis (outside the KQ75 region) reveals that the skewness of the Q, V, and W 
maps is -0.124, -0.125, and -0.127, respectively, 
indicating that the temperature distribution is skewed to the left with a longer tail for negative 
CMB values. 
For comparison, notice that Monte Carlos Gaussian maps (described in detail below) have 
skewness mean: $0.0012 \pm 0.0853$. 
Notice also that in less than 5\% of the Monte Carlos Gaussian maps one finds skewness 
values larger than those values found in WMAP data and this fact just represent the 
ergodicity of the data. 

For completeness, we shall analyze the temperature distribution in original and modified 
WMAP maps, termed X-cases, in which the cold-spot temperature is reduced. 
Consider the set of $n$ cold-spot pixels, that is, those pixels of the WMAP map belonging 
to the cap centered at $(209^{\circ},-57^{\circ})$ and within $8^{\circ}$ of radius. 
Let $\{ T_i^{\mbox{\bf\tiny CS}}\,;\, i=1,\ldots,n \}$ be the set of cold spot pixels' temperatures 
of a WMAP map: 
the X-case corresponds to reducing these temperatures to X\% of the original ones, 
while leaving all the other map pixels intact. 
For instance, X $\!$= $\!$90 means that the cold-spot pixels' temperatures were reduced 
to the values $0.9 \times T_i ^{\mbox{\bf\tiny CS}}$. 
Clearly, X=100 or 100\%-case refers to the original WMAP map. 
For X = 0 we replace the cold-spot temperature by the mean temperature of the 
WMAP map outside the KQ75 region, that is, $\sim -10^{-5\,}$mK. 

We exhibit the temperature distribution for the V map in the bottom panel of Fig.~\ref{fig1}, 
although in an atypical format in order to enhance possible deviations from a Gaussian 
distribution. 
In the horizontal axis we plot the square of the temperatures $T^2$ 
and in the vertical axis we have the number of pixels in logarithmic scale. 
In this picture, a Gaussian curve 
$= (1/\sigma_{\mbox{\sc\footnotesize g}} \sqrt{2\pi}) \, 
e^{-\frac{1}{2} (T^2 / \sigma_{\mbox{\sc\tiny G}}^2)}$ 
is transformed into a straight line. 
We plotted four data sets: 
(i) The straight (blue) line is the expected Gaussian distribution, 
for positive and negative CMB temperature fluctuations because a Gaussian 
is symmetric with respect to zero, 
with variance $\sigma_{\mbox{\sc\footnotesize g}}^2 = 4.65 \times 10^{-4}$ mK$^{2}$. 
(ii)~The histogram (red) curve corresponds to the negative temperatures of the V map. 
(iii) The dashed (violet) line corresponds to the positive temperatures of the V map. 
(iv) The triangles represent the negative temperatures of the case X=70 for the V map. 
As predicted by the skewness value, the (red) histogram curve reveals a non-linear tail 
due to pixels with anomalously large negative temperatures $T \lesssim -0.07$ mK 
(or $T^2 \gtrsim 0.0045$ mK$^{2}$), and we want to know if the cold spot~\cite{cold-spot} 
is related to this effect. 
Instead of masking the cold spot in a given WMAP map (which, at the angular scales 
we are interested, shall introduce a spurious signal in the next NS-asymmetry analysis) 
we investigate several X-cases. 
In fact, we discover that when the cold-spot temperature is 70\% of its original value, 
or lower, the non-linear tail disappears. 
Consistently, in the 70\%-case the skewness is -0.0323, -0.0294, and -0.0334, for 
the Q, V, and W maps, respectively. 
In conclusion, the diminution of the cold-spot temperature implies the suppression 
of the non-linear tail in the temperature distribution plot. 

\section{sigma-map method} \label{method} 

Now we are interested in studying the ef\/fect of these cold-spot temperature changes on 
the angular-correlation strengths in WMAP maps.  
For this we use a geometrical-statistical method that leads us to quantify, in intensity and 
direction, the CMB angular correlations, in particular, to search for an hemispherical asymmetry 
(for details of this method see Ref.~\cite{BFW}). 

Let $\Omega_{\gamma_0}^J \equiv \Omega(\theta_J,\phi_J;\gamma_0) 
\subset {\cal S}^2$ be a spherical cap region on the celestial sphere ${\cal S}^2$, 
of $\gamma_0$ degrees of aperture, with vertex at the $J$th pixel, 
$J = 1, \ldots, N_{\mbox{\footnotesize caps}}$, where $(\theta_J,\phi_J)$ 
are the angular coordinates of the $J$th pixel's center.
Both the number of spherical caps $N_{\mbox{\footnotesize caps}}$ and the coordinates 
of their centers $(\theta_J,\phi_J)$ are def\/ined using the {\sc healpix} pixelization 
scheme~\cite{Gorski}. 
The union of the $N_{\mbox{\footnotesize caps}}$ spherical caps covers completely 
${\cal S}^2$. 

Given a pixelized CMB map, the 2-point angular-correlation function (2PACF) of the 
temperature f\/luctuations, $T=T(\theta,\phi)$, corresponding to the pixels located in the 
spherical cap $\Omega_{\gamma_0}^J$ is def\/ined by 
$\mbox{\rm C}(\gamma)^J \equiv \langle\, T(\theta_i,\phi_i) T(\theta_{i'},\phi_{i'}) \,\rangle$, 
where 
$\cos\gamma = \cos\theta_i \cos\theta_{i'} 
+ \sin\theta_i \sin\theta_{i'} \cos(\phi_i\!-\!\phi_{i'})$, 
and $\gamma \in (0,2\gamma_0]$ is the angular distance between the $i$th and the 
$i'$th pixels' centers. 
The average $\langle \,\, \rangle$ in the above definition is done over all the products 
$T(\theta_i,\phi_i) T(\theta_{i'},\phi_{i'})$ such that 
$\gamma_k \equiv \gamma \in ((k-1)\delta,\, k\delta]$, for 
$k = 1,..., N_{\mbox{\footnotesize bins}}$, where 
$\delta \equiv 2\gamma_0 / N_{\mbox{\footnotesize bins}}$
is the bin width. 
We denote by $\mbox{\rm C}_k^J \equiv \mbox{\rm C}(\gamma_k)^J$ the value of the 
2PACF for the angular distances $\gamma_k \in ((k-1)\delta,\, k\delta]$. 
%
%
Now, consider a scalar function 
$\sigma: \Omega_{\gamma_0}^J \subset {\cal S}^2 \mapsto {\Re}^{+}$, for 
$J = 1, \ldots, N_{\mbox{\footnotesize caps}}$, which assigns to the $J$ cap, 
centered at $(\theta_J,\phi_J)$, a real positive number 
$\sigma_{J} \equiv \sigma(\theta_J,\phi_J) \in \Re^+$. 
We def\/ine $\sigma_{J}$ as 
\vspace{-0.33cm}
\begin{equation} \label{sigma}
\sigma^2_J  \equiv \frac{1}{N_{\mbox{\footnotesize bins}}}
\sum_{k=1}^{N_{\mbox{\footnotesize bins}}} (\mbox{\rm C}_k^J)^2 \,\, . 
\end{equation}
To obtain a quantitative measure of the angular correlations in a CMB map, we cover 
the celestial sphere with $N_{\mbox{\footnotesize caps}}$ spherical caps, and calculate 
the set of sigma values $\{ \sigma_{J}, \, J=1,...,N_{\mbox{\footnotesize caps}} \}$ using 
Eq.~(\ref{sigma}). 
Associating the sigma value $\sigma_J$ to the $J$th pixel, 
for $J=1, \ldots, N_{\mbox{\footnotesize caps}}$, 
one f\/ills the celestial sphere with positive real numbers. 
Then, according to a linear scale (where $\sigma^{\mbox{\footnotesize min}} \rightarrow blue$,  
$\sigma^{\mbox{\footnotesize max}} \rightarrow red$), one converts this sigma-values map 
into a colored map: this is the sigma map. 
F\/inally, we f\/ind the multipole components of a sigma map 
$\sigma(\theta,\phi) = \sum_{\ell,\, m} A_{\ell\, m} Y_{\ell\, m}(\theta,\phi)$, 
and calculate its angular power spectrum $\{ \mbox{\sc S}_{\ell},\, \ell=1,2,... \}$, that is  
\begin{eqnarray} \label{aps} 
\mbox{\sc S}_{\ell} \equiv \frac{1}{2\ell+1} \sum_{m={\mbox{\small -}}\ell}^{\ell} \, |A_{\ell\, m}|^2 \, . 
\end{eqnarray}

Accordingly, the power spectrum $\mbox{\sc S}_{\ell}$ of a sigma map computed from a WMAP 
map provides quantitative information of its statistical anisotropy features as compared with 
the mean of the sigma-map power spectra obtained from simulated isotropic Gaussian CMB 
maps. 
With this aim we produced a set of $1\,000$ Monte Carlo (MC) CMB maps, which correspond 
to random realizations seeded by the $\Lambda$CDM angular power spectra~\cite{wmap}, 
with $\ell = 2 - 512$ ($N_{\mbox{\footnotesize\rm side}}=256$). 
From these full-spectrum MC maps we generate, similarly as we have done with the WMAP data, 
a set of MC maps containing the multipoles $\ell = 11 - 20$ 
($N_{\mbox{\footnotesize\rm side}}=32$). 
We then calculate the sigma maps of these MC maps (hereafter sigma-maps MC) and their 
corresponding angular power spectra using eq.~(\ref{aps}). 
Last, we compute statistical significance by comparing the power spectra of the sigma-maps 
WMAP with the spectra from the sigma-maps MC.
For these spectra, $\mbox{\sc S}_{\ell} \simeq 0$, for $\ell \ge 5$, thus we only consider the 
multipole range $\{ \mbox{\sc S}_{\ell}, \,\ell=1,\!...,5 \}$. 
$\mbox{\sc S}_{1}$ corresponds to the dipolar anisotropy strength.

\begin{figure} 
\includegraphics[width=7.5cm,height=13cm]{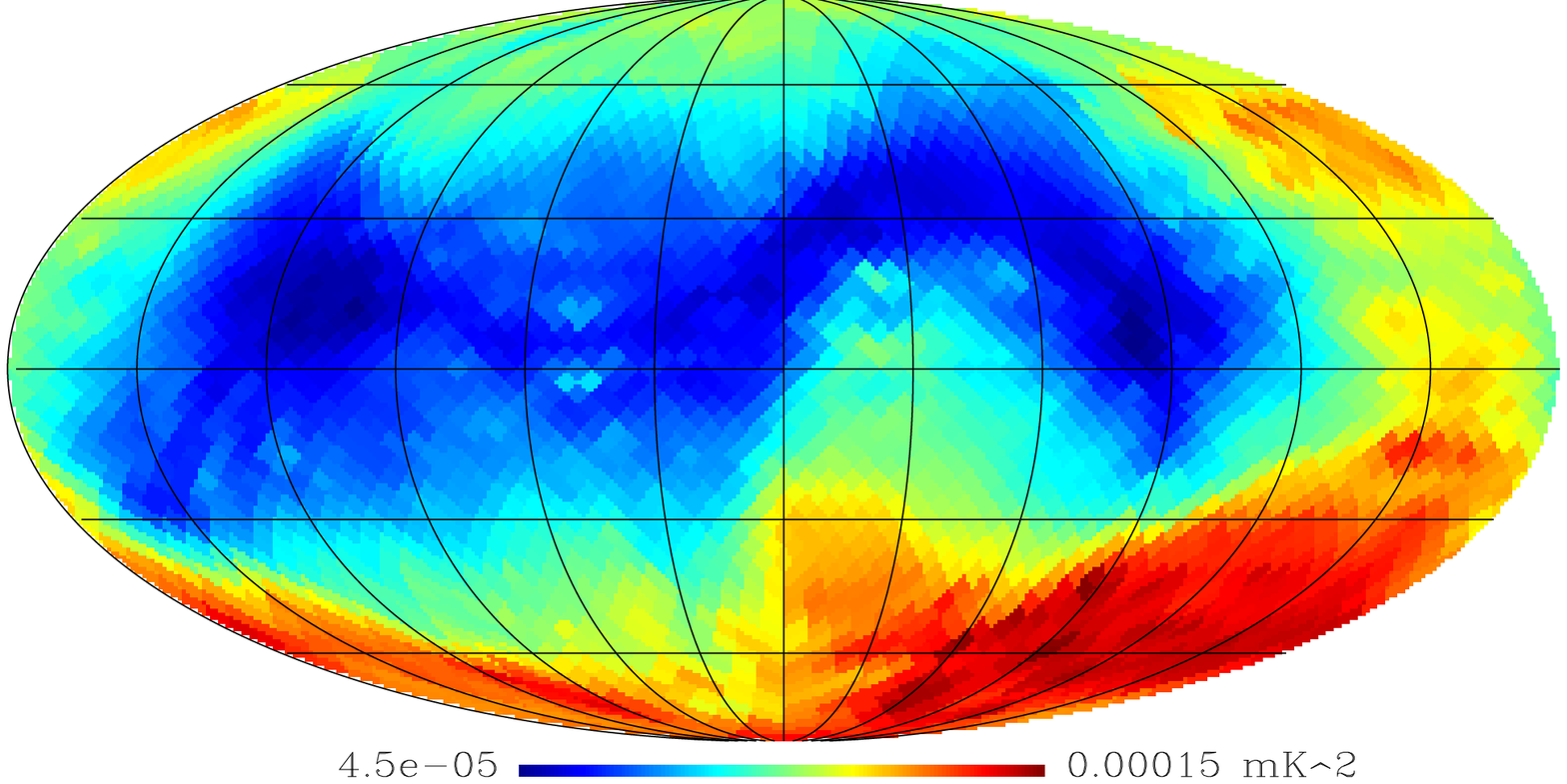}

\vspace{-7.5cm}
\mbox{\hspace{-0.6cm}
\includegraphics[width=9.3cm,height=14.cm]{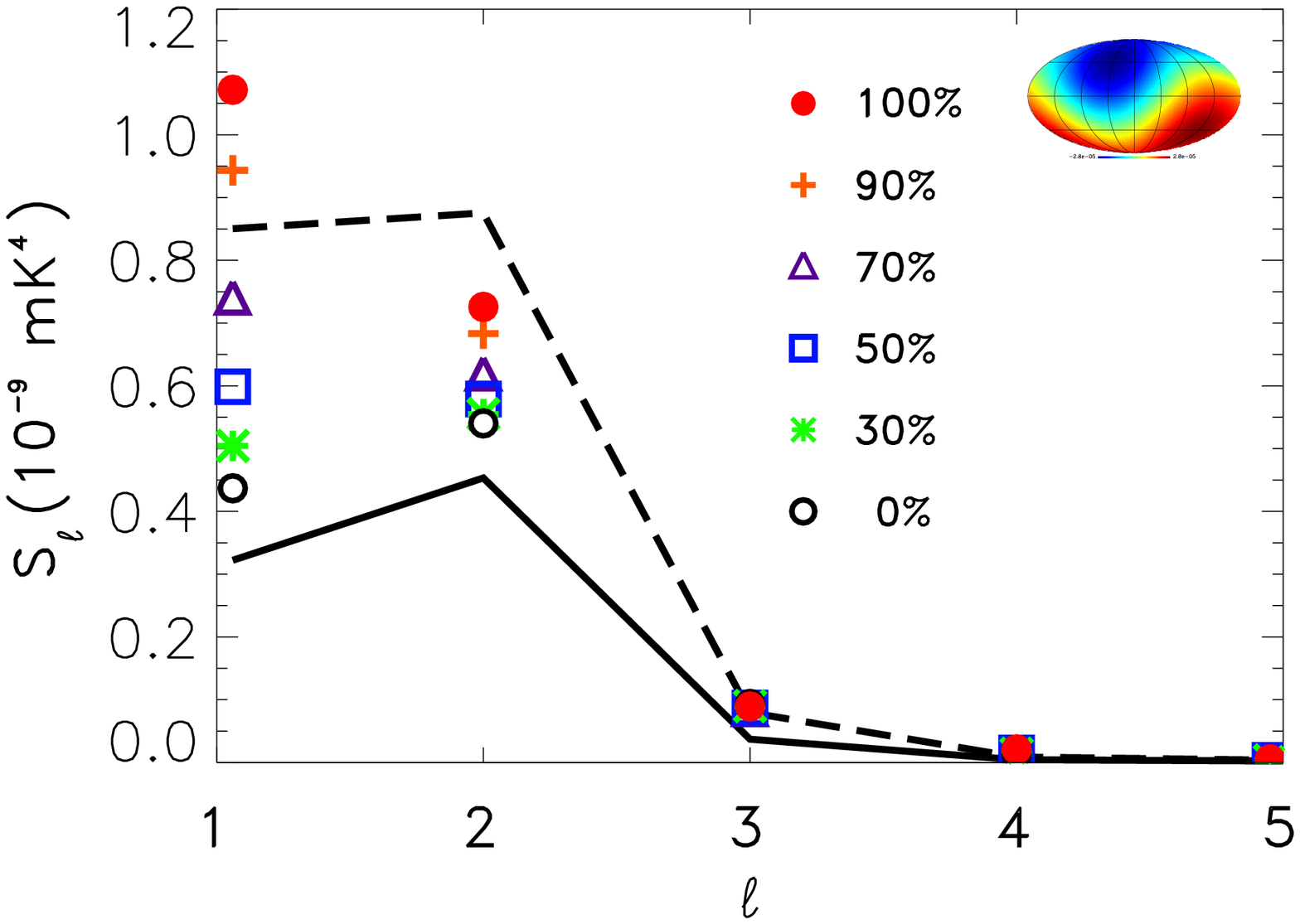}}
\vspace{-6.8cm}
\caption{\label{fig2}
Top: This is the sigma-map WMAP-V obtained using 
$\gamma_0 = 45^{\circ}$, $N_{\mbox{\footnotesize bins}}=45$, and 
$N_{\mbox{\footnotesize caps}}=3\,072$. 
The NS-asymmetry phenomenon is clearly seen in the uneven hemispherical distribution of 
the angular correlation's strengths, depicted as red and blue pixels (large and small $\sigma_J$, 
respectively). 
Bottom: 
Angular power spectra of the sigma map shown above (100\%-case) and sigma maps from 
several cases: X = 90, 70, 50, 30, and 0, where the cold-spot temperature in the V map is a 
fraction, X\%, of its original value.  
The solid (dashed) line corresponds to the mean (95.4\% CL) spectra of $1\,000$ sigma-maps 
MC.
The statistical significance of the NS-asymmetry is revealed at 97.6\% CL in the 100\%-case, 
while for the 70\%-case it is 91.5\% CL. 
The dipole component of the above sigma map, displayed in the inset, 
points toward $(l,b) \simeq (225^{\circ},-45^{\circ})$. 
} 
\end{figure}

\section{Results and Conclusions} \label{results and conclusions} 

As observed in Fig.~\ref{fig2} top's panel, this sigma-map WMAP-V clearly exhibits a dipolar 
red-blue region indicating a hemispherical asymmetry in the distribution of the angular 
correlations power, this is the NS-asymmetry anomaly at medium angular scales 
($\ell=11 - 20$). 
In fact, in the bottom panel of Fig.~\ref{fig2} we show its angular power spectrum and confirm, 
for the 100\%-case, that its dipole term is indeed anomalous, unexpected in 97.6\% of the 
sigma-maps MC for the sigma-map V (97.6\% and 97.5\% for sigma-map Q and sigma-map W, 
respectively). 
Moreover, we studied the effects caused by a systematic reduction of the cold-spot 
temperature. 
Our results, for the cases X = 90, 70, 50, 30, and 0, shown also in the bottom panel of 
Fig.~\ref{fig2}, illustrate that successively lowering the cold-spot temperature implies the 
monotonic decrease of the dipole intensity, $\mbox{\sc S}_{1}$, which in turn means the loss 
of statistical significance of the NS-asymmetry phenomenon. 
Consistently, the triangles data corresponding to the 70\%-case shown in Figs.~\ref{fig1} 
and~\ref{fig2}, illustrate the 
correlation found here: 
the weakening of non-Gaussian features in the cold spot 
implies the fading of the anomalous NS-asymmetry in WMAP maps. 
A correlation between a non-Gaussian feature (i.e., the cold spot) and a large-scale 
anisotropy (i.e., the NS-asymmetry) has been established. 
Another crucial issue concerns the case X=0, which corresponds to the absence of the 
cold spot in CMB maps.
For X=0 the dipole intensity $\mbox{\sc S}_{1}$ decreases 59.6\% with respect to the 
case X=100 for the sigma-map V (while it decreases 58.7\% and 60.4\% for the sigma-map 
Q and sigma-map W, respectively), achieving a value very close to the 
mean obtained averaging the sigma-maps-MC spectra. 
Additionally, we have investigated nine other spots, the coldest and hottest ones in the 
southern hemisphere of WMAP maps, to know their influence on NS-asymmetry 
phenomenon. 
Considering separately the case X=0 for these nine spots, our results show that none of 
them reduces the confidence level of the hemispherical asymmetry below 90\% C.L. 
In conclusion, the (non-Gaussian) cold spot 
is responsible for 60\% of the anomalous NS-asymmetry power observed in WMAP data, 
at medium angular scales.

\begin{table}
\caption{\label{table1} 
Robustness tests for sigma-map analyses using several parameters. 
Here are the confidence levels (CL) for $\mbox{\sc S}_{1}$ in the NS-asymmetry analyses 
in Q, V, and W maps ($N_{\mbox{\footnotesize side}}=32$, pixel size $1.8^{\circ} \!\Rightarrow$
$\delta \equiv 2\gamma_{_0} / N_{\mbox{\footnotesize bins}}$ 
 {\footnotesize$\gtrsim$} $2^{\circ}$).  
The C.L. intervals correspond to different binning tests: 
$N_{\mbox{\footnotesize bins}} = 15, 30, 45, 60$, provided 
$\gamma_{_0} \!\ge \!N_{\mbox{\footnotesize bins}}$ to minimize the statistical noise 
in $\sigma_{J}$ calculations.
}
\vspace{0.3cm}
\begin{tabular}{|l||c|c|c|c|} 
\hline
\hline  
maps {\large $\!\!\backslash$} 
$\!\!(N_{\mbox{\footnotesize caps}}, \gamma_{_0})$ 
& (768,$30^{\circ}$) & (768,$45^{\circ}$) & (3072,$45^{\circ}$) & (768,$60^{\circ}$)$\!$ \\ 
\hline
\,\,\,\,\,\,Q           & 94\%$-$96\%     & 97\%$-$98\%     & 97\%$-$98\%     & 94\%$-$96\%  \\
\,\,\,\,\,\,V           & 94\%$-$96\%     & 97\%$-$98\%     & 97\%$-$98\%     & 94\%$-$96\%  \\
\,\,\,\,$\:$W        & 94\%$-$96\%     & 97\%$-$98\%     & 97\%$-$98\%     & 94\%$-$96\%  \\
\hline
\hline
\end{tabular}
\end{table}

It is well known~\cite{wmap} that CMB foregrounds are frequency dependent. 
For this, in order to find out possible foreground signatures in our findings we used the Q, V, 
and W band maps. 
In fact, robustness tests (see Table~\ref{table1}) and sigma-map outcomes studying these 
three maps are not significantly different, meaning that our results are unlikely due to residual 
foregrounds. 
Pixel noise in WMAP maps is another possible source of incorrect results. 
However, at the angular scales we are working, pixel noise artifact is dominated by the 
cosmic variance~\cite{wmap}, and this effect was already included in Monte Carlo maps. 
In fact, pixel noise in WMAP data for $\ell=11 - 20$ is of the order 
$1\mu$K, while temperature uncertainty due to cosmic variance is $\sim 5\mu$K. 
Consequently, the robustness of our analyses supports the validity of our NS-asymmetry 
results. 

There is strong evidence for NS-asymmetry at several angular scales in WMAP 
data~\cite{NS-asymmetry}. 
In addition, we know that different angular scales in the CMB field encode information 
corresponding to distinct physical phenomena. 
Therefore, it is highly plausible that NS-asymmetry at different scales has not a common 
origin. 
One manifestation of this is the fact that the dipolar direction of hemispherical asymmetry 
is distinct for large ($\ell = 2 - 10$) and medium ($\ell = 11 - 20$) angular scales, i.e., 
$\sim(220^{\circ},-20^{\circ})$~\cite{NS-asymmetry} and $\sim(225^{\circ},-45^{\circ})$, 
respectively. 
Here we focused on those angular scales compatible with cold-spot dimensions, that is 
$\theta \simeq 9^{\circ} \!-\! 16^{\circ}$~\cite{cold-spot} 
(i.e., $\ell=11 - 20$, where $\ell \sim \pi / \theta$). 
Clearly, the effect of the cold spot does not starts at $\ell=11$ neither does it end at $\ell=20$. 
For $\ell \le 10$ the cold-spot influence contends with colder and hotter spots, this is because 
the angular power spectra $C_{\ell} \propto 1/(\ell (\ell+1))$ (Sachs-Wolfe effect) is larger 
for smaller $\ell$, and seems difficult to cause the NS-asymmetry at large-angles. 
On the other hand, the impact of the cold spot on NS-asymmetry for smaller scales, 
$\ell \ge 21$, is unknown and deserves further investigation. 
These facts lead us to conclude that it might be possible that distinct phenomena are 
causing the reported hemispherical asymmetry at several angular 
scales~\cite{NS-asymmetry}, and what is really happening is that 
different phenomena are predominant at distinct angular scales. 
Of course, the origin of the cold spot, including the possibility that it is just an anomalous 
statistical fluke, is still an open question.

\section*{Acknowledgements}
We thank A. F. F. Teixeira, M. J. Rebou\c{c}as, and G. D. Starkman for critical reading 
and suggestions. 
This work was supported by CNPq (309388/2008-2). 
We are grateful for the use of the Legacy Archive for Microwave Background 
Data Analysis (LAMBDA)~\cite{wmap}. 
Some of the results in this paper have been derived using the {\sc healpix} 
package~\cite{Gorski}. 

\vspace{-0.2cm}


\begin{thebibliography}{99}

\vspace{-0.3cm}
\bibitem{wmap} 
C. L. Bennett {\it et al.}, Astrophys. J. Suppl. Ser. \textbf{148}, 1 (2003); 
G. Hinshaw {\it et al.}, Astrophys. J. Suppl. Ser. \textbf{170}, 288 (2007); 
G. Hinshaw {\it et al.}, Astrophys. J. Suppl. Ser. \textbf{180}, 225 (2009).

\bibitem{NS-asymmetry}
F. K. Hansen, A. J. Banday, and K. M. G\'orski, 
Mon. Not. $\!$R. $\!$Astron. $\!$Soc. $\!$\textbf{354}, $\!$641 $\!$(2004); 
%
H. K. Eriksen {\it et al.}, 
\apj \textbf{605}, 14 (2004); 
%
K. Land and J. Magueijo, \prl \textbf{95}, 071301 (2005);
%
C. J. Copi {\it et al.}, 
Mon. Not. R. Astron. Soc. \textbf{367}, 79 (2006); 
%
A. Bernui {\it et al.}, 
Astron. Astrophys. \textbf{454}, 409 (2006);
%
Y. Wiaux, P. Vielva, E. Mart\'{\i}nez-Gonz\'alez, and P. Vandergheynst, 
\prl \textbf{96}, 151303 (2006);
%
L. R. Abramo, A. Bernui, I. S. Ferreira, T. Villela, and C. A. Wuensche, 
\prd \textbf{74}, 063506 (2006); 
%
C. J. Copi, D. Huterer, D. J. Schwarz, and G. D. Starkman, 
\prd \textbf{75}, 023507 (2007);
%
D. Huterer, New Astron. Rev. \textbf{50}, 868 (2006);
%
K. Land and J. Magueijo, 
Mon. Not. R. Astron. Soc. \textbf{378}, 153 (2007);
%
A. Bernui {\it et al.}, 
Astron. Astrophys. \textbf{464}, 479 (2007);
%
P. K. Samal {\it et al.}, 
arXiv:0708.2816 [astro-ph];
%
P. K. Samal {\it et al.}, 
Mon. Not. R. Astron. Soc. \textbf{396}, 511 (2009);
%
B. Lew, JCAP 0808:017 (2008); 
%
B. Lew, JCAP 0809:023 (2008);
%
A. Bernui and W. S. Hip\'olito-Ricaldi, Mon. Not. R. Astron. Soc. \textbf{389}, 1453 (2008); 
%
F. K. Hansen {\it et al.}, 
arXiv:0812.3795 [astro-ph];
%
J. Hoftuft {\it et al.}, 
\apj \textbf{699}, 985 (2009); 
%
Y. Ayaita, M. Weber, and C. Wetterich, 
arXiv:0905.3324 [astro-ph];
%
M. Frommert and T. A. Ensslin, arXiv:0908.0453 [astro-ph];
%
C. Dickinson {\it et al.}, arXiv:0903.4311 [astro-ph];
%
L. R. Abramo, A. Bernui, and T. S. Pereira, 
arXiv:0909.5395 [astro-ph].
%


\bibitem{cold-spot}
P. Vielva {\it et al.}, 
\apj \textbf{609}, 22 (2004);
%
M. Cruz {\it et al.}, 
Mon. Not. R. Astron. Soc. \textbf{356}, 29 (2005);
%
M. Cruz {\it et al.},
Mon. Not. R. Astron. Soc. \textbf{369}, 57 (2006);
%
P. Vielva {\it et al.},
Mon. Not. R. Astron. Soc. \textbf{381}, 932 (2007);
%
Y. Wiaux {\it et al.}, 
Mon. Not. R. Astron. Soc. \textbf{385}, 939 (2008);
%
J. D. McEwen {\it et al.}, 
Mon. Not. R. Astron. Soc. \textbf{ 388}, 659 (2008);
%
P. Naselsky {\it et al.}, 
arXiv:0712.1118 [astro-ph]; 
%
D. Pietrobon {\it et al.}, 
\prd \textbf{78}, 103504 (2008);
%
M. Cruz, E. Mart\'{\i}nez-Gonz\'alez and P. Vielva, 
arXiv:0901.1986 [astro-ph].
%
D. Pietrobon {\it et al.}, 
arXiv:0905.3702 [astro-ph];
%
G. Rossmanith {\it et al.}, 
arXiv:0905.2854 [astro-ph].


\bibitem{foregrounds}
%
P. D. Naselsky, L.-Y. Chiang, P. Olesen, and I. Novikov, 
\prd \textbf{72} (2005) 063512; 
%
A. de Oliveira-Costa and M. Tegmark, \prd \textbf{74}, 023005 (2006); 
%
A. Gruppuso and C. Burigana, 
JCAP 0908:004 (2009).
%
%
P. Bielewicz {\it et al.}, 
\apj \textbf{635}, 750 (2005); 
%
A. Bernui {\it et al.}, 
Int. Journal of Mod. Phys. D \textbf{16}, 411 (2007);
%
%
P. Vielva and J. L. Sanz, 
Mon. Not. R. Astron. Soc. \textbf{397}, 837 (2009);
%
M. Kawasaki {\it et al.},
JCAP 0901:042 (2009);
%
A. Bernui and M. J. Rebou\c{c}as, 
\prd \textbf{79}, 063528 (2009);
%
%
D. J. Schwarz, G. D. Starkman, D. Huterer, and C. J. Copi, 
\prl \textbf{93}, 221301 (2004); 
%
E. F. Bunn, \prd \textbf{75}, 083517 (2007);
%


\bibitem{mechanisms}
C. Gordon, W. Hu, D. Huterer, and T. Crawford, 
\prd \textbf{72}, 103002 (2005); 
%
L. Ackerman, S. M. Carroll, and M. B. Wise, \prd \textbf{75}, 083502 (2007);
%
T. S. Pereira, C. Pitrou, and J.-P. Uzan, JCAP 0709:006 (2007);
%
C. Pitrou, T. S. Pereira, and J.-P. Uzan, JCAP 0804:004 (2008);
%
T. S. Pereira and L. R. Abramo, 
\prd \textbf{80}, 063525 (2009);
%
L. Campanelli, P. Cea, and L. Tedesco, 
\prl \textbf{97}, 131302 (2006); 
%
L. Campanelli, P. Cea, and L. Tedesco, 
\prd \textbf{76}, 063007 (2007); 
%
L. Campanelli, 
\prd, \textbf{80}, 063006 (2009);
%
A. L. Erickcek, M. Kamionkowski, and S. M. Carroll, 
\prd \textbf{78}, 123520 (2008);
%
Y. Shtanov and H. Pyatkovska, 
Phys. Rev. D \textbf{80}, 023521 (2009); 
%
I. Y. Aref\'eva, N. V. Bulatov, L. V. Joukovskaya, and S. Y. Vernov, 
arXiv:0903.5264 [hep-th];
%
C. M. Hirata, 
JCAP 0909:011 (2009);
%
T. Kahniashvili, G. Lavrelashvili, and B. Ratra, \prd \textbf{78}, 063012 (2008);
%
T. R. Seshadri and K. Subramanian, 
\prl \textbf{103}, 081303 (2009);
%
C. Caprini, F. Finelli, D. Paoletti, and A. Riotto,  
JCAP  0906:021 (2009);
%
J. Kim and P. Naselsky, 
JCAP 0907:041 (2009). 
%


\bibitem{correl-anom1}
A. Raki\'c and D. J. Schwarz, \prd \textbf{75}, 103002 (2007).

\bibitem{correl-anom2}
A. Bernui, \prd \textbf{78}, 063531 (2008).

\bibitem{correl-anom3}
D. Pietrobon, arXiv:0907.4443 [astro-ph].

\bibitem{Gorski} 
K. M. G\'orski {\it et al.}, 
\apj \textbf{622}, 759 (2005).

\bibitem{BFW} A. Bernui, I. S. Ferreira, and C. A. Wuensche, 
\apj  \textbf{673},  968 (2008). 

\end{thebibliography}
\end{document}